\documentclass[12pt,draftcls,onecolumn]{IEEEtran}
\usepackage{amsmath}
\usepackage{amsthm}
\usepackage{amsfonts}
\usepackage{amssymb}
\usepackage[utf8]{inputenc}
\DeclareUnicodeCharacter{00A0}{ }
\renewcommand{\vec}[1]{\mathbf{#1}}
\usepackage[justification=centering]{caption}
\usepackage{cite}
\usepackage{graphicx}
\usepackage{xcolor}
\newcommand{\vast}{\bBigg@{4}}
\newcommand{\Vast}{\bBigg@{5}}

\usepackage[]{algorithm2e}
\RestyleAlgo{boxruled}
\LinesNumbered
\bstctlcite{IEEEexample:BSTcontrol}
\title{Max-Min Data Rate Optimization for RIS-aided Uplink  Communications with Green Constraints
	\author{Athira Subhash, Abla Kammoun, Ahmed Elzanaty, Sheetal Kalyani, Yazan H. Al-Badarneh, and Mohamed-Slim Alouini} \thanks{A. Subhash, and S. Kalyani are with the Department of Electrical Engineering, Indian Institute of Technology, Madras, India. (email:\{ee16d027@smail,skalyani@ee\}.iitm.ac.in).\\A. Elzanaty is with the 5GIC \& 6GIC, Institute for Communication Systems (ICS), University of Surrey, Guildford, GU2 7XH, United Kingdom (e-mail: a.elzanaty@surrey.ac.uk)
\\Y. H. Al-Badarneh is with the Department of Electrical Engineering, The University of Jordan, Amman, 11942 (email: yalbadarneh@ju.edu.jo). \\ A. Kammoun and M.-S. Alouini are with the Computer Electrical and Mathematical Sciences and Engineering (CEMSE) Division, King Abdullah University of Science and Technology (KAUST), Thuwal, Makkah Province, Saudi Arabia. (e-mail:\{abla.kammoun,slim.alouini\}@kaust.edu.sa).}}
	
\begin{document}
\maketitle
\begin{abstract}
Smart radio environments aided by reconfigurable intelligent reflecting surfaces (RIS) have attracted much research attention recently. We propose a joint optimization strategy for beamforming, RIS phases, and power allocation to maximize the minimum SINR of an uplink RIS-aided communication system. The users are subject to constraints on their transmit power. {We derive a closed-form expression for the beam forming vectors and a geometric programming-based solution for power allocation. We also propose two solutions for optimizing the phase shifts at the RIS, one based on the matrix lifting method and one using an approximation for the minimum function. We also propose a heuristic algorithm for optimizing quantized phase shift values.} The proposed algorithms are of practical interest for systems with constraints on the maximum allowable electromagnetic field exposure. {For instance,  considering  $24$-element RIS, $12$-antenna BS, and $6$ users,  numerical results show that the proposed algorithm achieves close to $300 \%$ gain in terms of minimum SINR compared to a scheme with random RIS phases.}
\end{abstract}

\section{Introduction}
The idea of a smart radio environment (SRE) redefines wireless environment as an optimization variable which can not only be adapted to but can be programmed and controlled along with the transmitter and receiver to improve communication quality \cite{di2020smart}. A reconfigurable intelligent reflecting surface (RIS) can passively modify the phase of the impinging signals and hence can be used to steer the received signal in the direction of interest.
Hence, it  offers considerable improvement in the performance of communication systems    \cite{di2020smart,basar2019,kammoun2020asymptotic}.
The RIS adds more degrees of freedom to the communication channel. Hence, 
the joint optimization of the beamforming (BF) strategy at the transmitter/receiver, transmit power allocation, and RIS phase shift design is essential to maximize the system performance. Some resource allocation schemes have been studied in the literature \cite{lv2021multiuser,xiu2021uplink,you2021reconfigurable}. {The authors of \cite{lv2021multiuser} propose an iterative algorithm for the joint resource allocation in a RIS-aided uplink network to minimize the mean square error.} The problem of maximizing the achievable rate under hardware constraints by jointly optimizing the ADC quantization bits, the
RIS phase shifts and the beam selection matrix is studied in \cite{xiu2021uplink}. The authors of \cite{you2021reconfigurable} propose an optimization framework for jointly
designing the transmit covariance matrices of the users and the
RIS phase shift matrix to maximize the system's global energy
efficiency. Assuming that the RIS phase shifts cannot take continuous values due to hardware constraints,  the authors of \cite{wu2019beamforming} model the RIS phases as discrete values. They propose an algorithm to jointly optimize the continuous
transmit precoder at the access point (AP) and discrete phase shifts at the RIS to minimize the transmit power at the AP. {A joint resource allocation strategy to minimize the symbol error rate in the uplink of a RIS-aided MIMO system with spatially modulated users is studied in \cite{luo2021spatial}.} All of these works and several closely related literature consider objective functions which do not ensure fair service to all the users. 
To the best of our knowledge, such a study ensuring fair power allocation and beamforming in the uplink with per-user power constraints is not addressed in the literature.
\par In this work, we study the joint optimization of the receive BF vector at the BS, transmit power of the users and the phase shifts at the RIS that maximize the minimum SINR subject to constraints on the transmit power of the users. We propose three solutions for the optimal phase shift design, two of which assume the RIS phase shift to be continuous and one assuming that the phase shifts are quantized. In a practical system with hardware limitations, one might not be able to realize any arbitrary phase shift value at the RIS and in such scenarios quantizing the optimal solution obtained by assuming the phase shift to be continuous might result in performance degradation. The proposed solution assuming discrete values for the phase shifts will be useful in such scenarios. Note that the minimum SINR achieved by assuming continuous values for the phase shifts will give the upper bound for the achievable system performance in any practical scenario and hence is of practical interest in deriving meaningful inferences from the system. \par We also demonstrate how the proposed results can handle additional green constraints like constraints on the users' maximum electromagnetic exposure (EMF). We call these constraints green because they ensure minimal resource usage for environment-aware scenarios. Recently there has been a rising concern over the exposure to radio frequency emissions from communication devices \cite{chiaraviglio2021health,sambo2014survey}. This has motivated many researchers to include metrics involving EMF in resource allocation problems \cite{ibraiwish2021emf,lou2021green,zappone2021optimization}. While \cite{ibraiwish2021emf} and \cite{zappone2021optimization} study the optimal resource allocation strategies that minimize exposure, the authors of \cite{lou2021green} introduce a constraint on the maximum exposure of the end-users. In this work, we also include constraints on the maximum exposure of each user.\\ The notations used in our letter are as follows: $\mathbb{C}^{p\times q}$ represents a complex matrix of dimensions $p$ by $q$, $\mathcal{CN}(\mu,\sigma^2)$ represents the complex normal random variable (RV) with mean $\mu$ and variance $\sigma^2$, $\text{diag}\left( a_1,\cdots,a_N\right)$ represents a diagonal matrix with diagonal elements $a_1,\cdots,a_N$, and $||\vec{a}||$ represents the $\ell_2$ norm of the vector $\vec{a}$. 
\color{black}
\section{System Model}
We consider the uplink communication of a multi user single input multiple output (SIMO) system where $K$ single-antenna users communicating with an $M$-antenna BS in the presence of an $N$-element RIS. The signal received at the BS is given by 
\begin{equation}
\vec{y} = \sum\limits_{k=1}^K \vec{H}_1 \mathbf{R}_{\textrm{R I S}}^{\frac{1}{2}} \vec{\Phi} \vec{h}_{2,k} x_k + \vec{n}
\end{equation}
where $\vec{H}_1 \in \mathbb{C}^{M \times N}$ represents the LOS channel matrix between the $\mathrm{BS}$ and $\mathrm{RIS}, \mathbf{R}_{\textrm{R I S}} \in \mathbb{C}^{N \times N}$
represents the spatial correlation matrix of the RIS elements, $\vec{H}_2 \in \mathbb
{C}^{N \times K}$ represents the channel between the $K$ users and the RIS, and the columns of $\vec{H}_2$ are given by $\mathbf{{h}}_{2, k}=\ell_k \mathbf{\tilde{h}}_{2, k}$ where $\ell_k$ is the path loss corresponding to the link between user $k$ and the RIS and $ \mathbf{\tilde{h}}_{2, k} \sim \mathcal{C} \mathcal{N}\left(\mathbf{0}, \mathbf{I}_{N}\right) \in \mathbb{C}^{N \times 1}$ denotes the fading channel gain between the RIS and user $k$, $\Vec{x}$ represents the transmit signal vector, and $\vec{n}$ represents the thermal noise, modeled as $\mathcal{C} \mathcal{N}\left(\boldsymbol{0}, \sigma^{2} \vec{I}_M\right)$. Furthermore, $\vec{\Phi}=\alpha\operatorname{diag}\left( \phi_{1}, \ldots, \phi_{N}\right) \in$ $\mathbb{C}^{N \times N}$ is the diagonal matrix accounting for the response of the RIS elements, $\alpha \in (0,1]$ is the fixed amplitude reflection coefficient and $\phi_{n}:=\exp \left(j \theta_{n}\right),n=1, \ldots, N$, where $\theta_n$ is the phase shift induced by the $n$-th RIS element. Now, let $\boldsymbol {\beta} \in \mathbb{C}^{K \times M }$ be the BF matrix at the BS and the received symbols can be decoded as $  \vec{r} = \boldsymbol {\beta} \vec{y}$. Then, the decoded symbol from user $k$ can be represented as $r_k = \boldsymbol{\beta}_k^H \vec{y}$,
where $\boldsymbol {\beta}_k^T$ is the $k$-th row of the matrix $\boldsymbol{\beta}$. The SINR for user $k$ at the BS is hence given by 
$$\text{SINR}_k = \frac{p_k\left|\boldsymbol {\beta}_k^H\vec{H}_1 \mathbf{R}_{\textrm{R I S}}^{\frac{1}{2}} \vec{\Phi} \vec{h}_{2,k} \right|^2}{\sum\limits_{i=1,i \neq k}^K p_i \left|\boldsymbol {\beta}_k^H\vec{H}_1 \mathbf{R}_{\textrm{R I S}}^{\frac{1}{2}} \vec{\Phi} \vec{h}_{2,i} \right|^2 + \sigma^2 ||\boldsymbol {\beta}_k||^2}.$$ The SINR experienced by each user is thus a function of the power allocation, BF at the BS, and the phase shift induced by the RIS elements. To ensure a fair service for all users, we would like to jointly optimize these variables such that the minimum SINR in the system is maximized, i.e., we would like to maximize $\min\limits_{k \in \mathcal{K}} \text{SINR}_k$. In the next section, we formulate an optimization problem to identify the optimal power allocation vector, the BF vectors at the BS and the phase shifts introduced by the IRS elements such that the constraints on the maximum transmit power of each user are satisfied.

\section{Optimization Problem and solution}

The optimization problem discussed in the last section can be mathematically formulated as follows:

\begin{equation}
\tag{$\mathcal{P}_1$}
\begin{aligned}
\max_{\{p_k\},\vec{\phi}, \vec{\beta}} \quad & \min\limits_k \ \text{SINR}_k \\
\textrm{s.t.} \quad &0 \leq  p_k \leq p_{\text{max},k};  \ k \in \{1,\cdots,K\},\label{prob1}\\
& ||\boldsymbol{\beta}_k||=1; \ k \in \{1,\cdots,K\},\\
&|\phi_n| =1,  \ n \in \{1,\cdots,N\}.
\end{aligned}
\end{equation}
Note that we have also imposed constraints on the norm of each BF vector and the magnitudes of the phase shift at each of the RIS elements. Next, we look at the solution for the above optimization problem assuming that the exact channel state information (CSI) is available at all nodes in the network. Note that this assumption might not be valid in all practical scenarios, and the SINR achieved, in this case, will result in an upper bound on the achievable system performance.
\subsection{Proposed solution}
The joint optimization problem of BF design, power allocation and phase shift design is not convex. Hence, it is not easy to arrive at a solution that jointly optimizes these three vectors.{Therefore, we use the popular technique of alternating optimization and solve three sub-problems iteratively, each optimizing one set of variables at a time.} We discuss the solution for each sub-problems in detail below. 
\subsection{BF vector design} \label{bf_opti}

In this sub-section, we solve the problem in $\mathcal{P}_1$ for a given choice of $\{p_k\}$ and $\boldsymbol{\Phi}$. Let $\vec{g}_k := \vec{H}_1 \mathbf{R}_{\textrm{R I S}}^{\frac{1}{2}} \vec{\Phi} \vec{h}_{2,k}$ and hence we have 
$$\text{SINR}_k = \frac{p_k\boldsymbol{\beta}_k^H \vec{g}_k \vec{g}_k^H \boldsymbol{\beta}_k} { \sum\limits_{i=1,i\neq k}^K p_i \boldsymbol{\beta}_k^H \vec{g}_i \vec{g}_i^H \boldsymbol{\beta}_k + \sigma^2 \boldsymbol{\beta}_k^H\boldsymbol{\beta}_k}.$$
The denominator of $\text{SINR}_k $ can be rewritten as 
$ \boldsymbol{\beta}_k^H \left( \sum\limits_{i=1,i\neq k} p_i \vec{g}_i \vec{g}_i^H + \sigma^2 \vec{I} \right)\boldsymbol{\beta}_k=\boldsymbol{\beta}_k^H \left( \boldsymbol{\Sigma}_k + \sigma^2 \vec{I} \right)\boldsymbol{\beta}_k$, where $\boldsymbol{\Sigma}_k = \sum\limits_{i=1,i\neq k} p_i \vec{g}_i \vec{g}_i^H$. 
Let $\boldsymbol{\mu}_k:= \left( \boldsymbol{\Sigma}_k + \sigma^2 \vec{I} \right)^{\frac{1}{2}} \boldsymbol{\beta}_k$ and hence we have, $\boldsymbol{\beta}_k =\left( \boldsymbol{\Sigma}_k + \sigma^2 \vec{I} \right)^{\frac{-1}{2}} \boldsymbol{\mu}_k $.
Thus, $$\text{SINR}_k= \frac{p_k\boldsymbol{\mu}_k^H\left( \boldsymbol{\Sigma}_k + \sigma^2 \vec{I} \right)^{\frac{-1}{2}} \vec{g}_k \vec{g}_k^H \left( \boldsymbol{\Sigma}_k + \sigma^2 \vec{I} \right)^{\frac{-1}{2}} \boldsymbol{\mu}_k }{\boldsymbol{\mu}_k^H \boldsymbol{\mu}_k} = \frac{p_k \boldsymbol{\mu}_k^H \vec{a}_k \vec{a}_k^H \boldsymbol{\mu}_k}{\boldsymbol{\mu}_k^H \boldsymbol{\mu}_k}$$ where $\vec{a}_k=\left( \boldsymbol{\Sigma}_k + \sigma^2 \vec{I} \right)^{\frac{-1}{2}} \vec{g}_k$. Note that the SINR of the $k$-th user will be maximized when $\boldsymbol{\mu}_k$ and $\vec{a}_k$ are both in the same direction. 
For $\boldsymbol{\mu}_k=\vec{a}_k$, we have 
\begin{equation}
\boldsymbol{\beta}_k = \frac{\left( \boldsymbol{\Sigma}_k + \sigma^2 \vec{I} \right)^{-1} \vec{g}_k}{\left\vert\left\vert{\left( \boldsymbol{\Sigma}_k + \sigma^2 \vec{I} \right)^{-1} \vec{g}_k}\right \vert\right \vert}.
\label{beta_opti}
\end{equation}
Furthermore, the SINR is now given by $\text{SINR}_k = p_k \vec{g}_k^H \left( \boldsymbol{\Sigma}_k + \sigma^2 \vec{I} \right)^{-1} \vec{g}_k$. Thus, for a given choice of phase vector and transmit power levels, the optimal receive BF for the $k$-th user's signal is given by (\ref{beta_opti}). 
\subsection{Power control} \label{power_opti}
Next, we find the optimal power allocation vector for a given choice of BF vectors and RIS phase shift values. Following steps similar to \cite[(2)]{cai2011unified}, this problem can be rewritten as 
\begin{align}
\min_{ \{p_k\},\tau} &  & \tau^{-1}  & & \\
\text { s.t. } & & \frac{\tau \left(\sum\limits_{i=1,i \neq k}^K p_i f_{k,i} + n_k \right)}{p_k f_{k,k}}  \leq 1  \ \ & \forall k \in\{1, \cdots, K\}, \nonumber \\
& & 0 \leq  p_k \leq p_{{\max},k}    \ \ & \forall k \in\{1, \cdots, K\},\nonumber 
\end{align} 
where $f_{k,i}=\left|\boldsymbol{\beta}_k^H\vec{H}_1 \mathbf{R}_{\textrm{R I S}}^{\frac{1}{2}} \vec{\Phi} \vec{h}_{2,i} \right|^2$ and $n_k=\sigma^2 $. 
The above problem can be solved using the popular technique of geometric programming. Geometric programming solvers are available in popular optimization packages like CVX and can be solved efficiently \cite{cvx}.

\subsection{Phase matrix design} \label{phase_opti}
In this sub-section, we study the problem of optimal phase matrix design for a given choice of BF vectors and power allocation. 

\subsubsection{Phase shift design for max-min SINR}
The problem is formulated as follows: 
\begin{align} \label{prob3}
\tag{$\mathcal{SP}_3$}
\max_{ \{\phi_n\}} & & \min_{k} \  \frac{p_k\left|\vec{b}_k^H \vec{\Phi}_{vec}  \right|^2}{\sum\limits_{i=1,i \neq k}^K p_i \left|\vec{b}_i^H \vec{\Phi}_{vec} \right|^2 + \tilde{\sigma}_k^2 } & & \nonumber \\
\text { s.t. } & &  |\phi_n| =1, \ \forall \ n \in \{1,\cdots,N\}, \nonumber 
\end{align}
where $\vec{b}_k^H=\boldsymbol{\beta}_k^H \vec{H}_1 \mathbf{R}_{\textrm{R I S}}^{\frac{1}{2}}\text{diag}(\vec{h}_{2,k})$, $\vec{\Phi}_{vec} = \alpha \left[\phi_1,\cdots,\phi_N \right]^T$ and $\tilde{\sigma}_k^2 =\sigma^2 ||\boldsymbol{\beta}_k||^2$. $\text{SINR}_k$ in the above expression can be rewritten as $$\text{SINR}_k=\frac{p_k\vec{\Phi}_{vec}^H \vec{R}_k \vec{\Phi}_{vec} }{\sum\limits_{i=1,i \neq k}^K p_i \vec{\Phi}_{vec}^H \vec{R}_i \vec{\Phi}_{vec}  + \tilde{\sigma}_k^2 }.$$ where $\vec{R}_i=\vec{b}_i\vec{b}_i^H$. We proceed by using the matrix lifting technique, similar to the authors of \cite{kafizov2021wireless}. Hence, we rewrite the problem in terms of the matrix $\vec{V}={\boldsymbol{\Phi}_{vec}} {\boldsymbol{\Phi}_{vec}}^{H}$ with the constraints ${\mathbf{V}} \succeq 0$ and $\operatorname{rank}({\mathbf{V}})=1$. We use semidefinite relation to relax the rank constraint and later use the Dinkelbach algorithm \cite[Algorithm 6]{zappone2015energy}  to solve the resulting problem. Finally, Gaussian randomization is used to arrive at a solution for $\boldsymbol{\Phi}_{vec}$, if the optimal $\vec{V}$ is not a rank one matrix. 

\subsubsection{Phase shift design using an approximation for the minimum}
Recall that for the optimal beamformer, the SINR of each user can be expressed as $\text{SINR}_k = {p}_k^* \vec{g}_k^H \left( \tilde{\Sigma}_k + \tilde{\sigma}^2 \vec{I} \right)^{-1} \vec{g}_k$. Next, we make use of a smooth approximation for the minimum function using the log sum exponential upper bound for the maximum. We have, $\max \{x_1,\cdots,x_n\} \leq  \log\left(e^{x_1}+e^{x_2}+\cdots + e^{x_n} \right)$. Given that each of $x_1,\cdots,x_n$ are non-negative, we have the following smooth approximation for the minimum function, $\min \left(x_1,\cdots,x_n \right) \geq \left({\log\left(e^{\frac{1}{x_1} }+e^{\frac{1}{x_2} }+\cdots + e^{\frac{1}{x_n} } \right)}\right)^{-1}$. 
Using this approximation, the optimization problem of interest can be written as:
$$\min_{ \{\phi_n\}} \  \text{OB}:={\log \left( \sum\limits_{k=1}^K e^{  \frac{1}{\rho_k}} \right)}  \ 
\text { s.t. } \  |\phi_n| =1, \ \forall \ n \in \{1,\cdots,N\}$$
where $\rho_k := {p}_k^* \vec{g}_k^H \left( {\Sigma}_k + {\sigma}^2 \vec{I} \right)^{-1} \vec{g}_k$. 

Given that the objective is a smooth function, we can now use the projected gradient descent to arrive at a solution for $\{\phi_n\}$ that minimizes the objective function of interest. The derivative of the objective function required for the algorithm is given by 
\begin{equation}
\frac{\partial \ \text{OB}}{\partial \phi_n}   =\frac{-\sum\limits_{k=1}^K \frac{e^{-\rho_k}}{\rho_k^2} \frac{\partial \rho_k}{\partial \phi_n}}{\sum\limits_{k=1}^K \exp(-\rho_k)}
\end{equation}
where $\frac{\partial \rho_k}{\partial \phi_n}  = 2 p_k^* \left\lbrace \left[\left(\vec{H}_1 \mathbf{R}_{\textrm{R I S}}^{\frac{1}{2}}   \right)^H  T_2   \vec{H}_1 \mathbf{R}_{\textrm{R I S}}^{\frac{1}{2}} \vec{\Phi} \vec{V}_k\right]_{n,n}- \right.$ $\left. \sum\limits_{i\neq k}^K {p}_i^* \left[ T_3^{2,i}T_2 T_1 T_2T_3^{1,i} \right]_{n,n} \right \rbrace . $
The detailed steps to derive the derivative of $\rho_k$ is given in Appendix \ref{d_rhok_by_phin_appendix}. 
\subsubsection{Quantized phase shift design}
This section explores the prospects of finding the optimal phase shift design after assuming that the phase shift values are quantized. Given $B$ bits representing each RIS element's phase, we will have $q:=2^B$ possible phase values. Here we propose a heuristic approach by searching over only a subset of the possible choices. We begin with an arbitrary initialization of the RIS phase vector. Then, we randomly choose an index of the RIS phase element and try swapping the current phase value with all the other possible values. A swap is retained if it results in an increase in the minimum SINR and is otherwise replaced by with previous value. The selection of random phase indices is continued until the last $L$ iterations of the algorithm result in almost the same SINR. This condition is expressed as $\sum\limits_{i=\text{index}-L-1}^{\text{index}-1} \left( \boldsymbol{\tau}_{\text{vec}}(i)-\boldsymbol{\tau}_{\text{vec}}(\text{index}) \right)< \epsilon$. As the value of $L$ increases, the time complexity increases. At the same time, we explore more possible choices for the phase vector and hence improve the minimum SINR. Therefore, the optimal choice of $L$ is to be decided based on the trade-off between exploration and computational complexity. The steps of the proposed heuristic are presented in Algorithm \ref{algo_phase_quant_new_rand}.


\section{Application}

This section presents one interesting application where the proposed resource allocation policy can be utilized. Recently, users' exposure to electromagnetic filed (EMF) has been a parameter of concern in the fifth-generation communication systems. The problem of optimal resource allocation that minimizes EMF of users while maintaining the quality of service (QoS) requirements has received much attention \cite{ibraiwish2021emf,lou2021green,zappone2021optimization,jiang2022hybrid}. Hence, we would like to ensure that each user's EMF is within a permissible limit while maximizing the minimum SINR. The EMF of the $k$-th user transmitting with a power $p_k$ can be evaluated as $\text{SAR}_{\text{ref},k}\times p_k$ where $\text{SAR}_{\text{ref},k}$ is the specific absorption rate (SAR) for user $k$ when the transmitted power is unity. Hence, we would like to introduce the following additional constraint in $\mathcal{P}_1$: $0 \leq  \text{SAR}_{\text{ref},k}p_k \leq \text{EMF}_{\text{max},k}$, where $\text{EMF}_{\text{max},k}$ is the maximum allowed exposure for user $k$ and $p_{\text{max}}$ is the maximum transmit power for each user. This new constraint can be combined with the existing power constraint in $\mathcal{P}_1$ and can be expressed as $0\leq p_k \leq \min \left \lbrace p_{\text{max}},\frac{\text{EMF}_{\text{max},k}}{\text{SAR}_{\text{ref},k}}\right \rbrace $. Hence, choosing $p_{\text{max},k} =\min \left \lbrace p_{\text{max}},\frac{\text{EMF}_{\text{max},k}}{\text{SAR}_{\text{ref},k}}\right \rbrace $, the joint optimization with EMF constraints can be solved using the method proposed in the previous section. In the next section, we present simulation results for this application and hence demonstrate the performance of the proposed solution methodology.

\begin{algorithm}[t!]
 \KwData{$B:$ number of bits, $q = 2^B$ number of phase levels, ${\theta}_i = \frac{2\pi (i-1)}{q}$, $i\in \{1,\cdots,q\}$. }
 \KwResult{$\tilde{\Vec{\Phi}}_{vec}$}
 $\Vec{\Phi}_{vec}^{0}$: initialization vector, $\boldsymbol{\tau}\left( \Vec{\Phi}_{vec}^{0} \right)$: SINR evaluated using $\Vec{\Phi}_{vec}^{0}$, $i=1$, $index=1$, $\boldsymbol{\tau}_{\text{vec}}(\text{index})=\boldsymbol{\tau}\left( \Vec{\Phi}_{vec}^{0} \right)$\;
 \While{$\left(\text{index}<L\Big|\Big | \sum\limits_{i=\text{index}-L-1}^{\text{index}-1} \left(\boldsymbol{\tau}_{\text{vec}}(i)- \boldsymbol{\tau}_{\text{vec}}(\text{index})\right)> \epsilon\right)$}{
  $\tilde{\Vec{\Phi}}_{vec}=\Vec{\Phi}_{vec}^{0}$\;
 Choose $n\in \{1,\cdots,N\}$ randomly\;
  \While{$i<N+1$}{
   $i=i+1$\;
 $\tilde{\Vec{\Phi}}_{vec}(n)=\exp \left( j \theta_i \right) $\;
 \textbf{if}($\boldsymbol{\tau}(\tilde{\Vec{\Phi}}_{vec})>\boldsymbol{\tau}(\Vec{\Phi}_{vec}^{0})$)\;
 $\Vec{\Phi}_{vec}^{0} =\tilde{\Vec{\Phi}}_{vec} $\;
 $\textbf{else}$ $\tilde{\Vec{\Phi}}_{vec}(n)=\exp \left( j \theta_{i-1} \right) $\;
index $=$ index $+1$ \;
 $\boldsymbol{\tau}_{\text{vec}}(\text{index})=\boldsymbol{\tau}({\Vec{\Phi}}_{vec}^0)$\;
   }
 }
 \caption{Randomized Heuristic for phase optimization}
 \label{algo_phase_quant_new_rand}
\end{algorithm}

\section{Simulation Results}
We consider a simulation area where the BS is located at the origin and the RIS is located at a point $\left(0.5,0.5\right)$ metres. The users are uniformly spread between the area of overlap between the first quadrant and two circles of radius $R_{min}=10 \ m$ and $R_{max}=70 \ m$ representing the exclusion zone and cell coverage, respectively. The channel between the BS and RIS and the IRS and the $k$-th user are given by $\Vec{H}_1 = \sqrt{\frac{\mathrm{PL}_{\mathrm{LOS}}\left(d^{\mathrm{RB}}\right)}{ N}} \left(\sqrt{\frac{\kappa}{\kappa+1} } \bar{\Vec{H}}_1+\sqrt{\frac{1}{\kappa+1}} \tilde{\mathbf{H}}_1\right) $ and $\mathbf{h}_{k}^{\mathrm{u}}=\sqrt{\operatorname{PL}_{\mathrm{NLOS}}\left(d_{k}^{\mathrm{UR}}\right)} \tilde{h}_{k}^{\mathrm{u}}$ respectively. Here, $\operatorname{PL}_{\mathrm{LOS}}\left(d\right)$ and $\operatorname{PL}_{\mathrm{NLOS}}\left(d\right)$ are the path losses experienced by the LOS and NLOS links at a distance of $d$ from the transmitter and are modeled using the 3GPP Urban
Micro (UMi) scenario from \cite[Table B.1.2.1-1]{3GPPTR36p814}. Throughout the simulations, they are chosen as  $\operatorname{PL}_{\mathrm{LOS}}(d) = \frac{10^\frac{G_{t}+G_{r}-35.95}{10}}{d^{2.2}}$ and $\operatorname{PL}_{\mathrm{NLOS}}(d) = \frac{10^\frac{G_{t}+G_{r}-33.05}{10}}{d^{3.67}}$, where $G_t$ and $G_r$ denote the antenna gains (in dBi) at the transmitter and receiver respectively. It is assumed that the elements of BS have $5$ dBi gain while each user antenna and the RIS elements have $0$ dBi gain and, $\kappa$ is the Rician factor of the link between the BS and the RIS and is chosen to be $10$ throughout the simulations. The elements of the $M \times N$ matrix $\bar{\Vec{H}}_1$ is given by,  $\left[\bar{\mathbf{H}}_{1}\right]_{m, n}  =  $ $  \exp \left[\frac{j2\pi}{\lambda} \left((m-1) d_{\text{BS}} \sin \left(\theta_{\text{LoS},1}(n)\right) \sin \left(\phi_{\text{LoS},1}(n)\right) + \right. \right.$ $\left.\left.(n-1) d_{\text{RIS}} \sin \left(\theta_{\text{LoS},2}(m)\right) \sin \left(\phi_{\text{LoS},2}(m)\right)\right) \right]$ where $\theta_{\text{LoS},i}(n)$ and $\phi_{\text{LoS},i}(n)$ for $i \in \{1,2\}$
and $n=1,\cdots,N$, are generated uniformly between $0$ to $\pi$ and $0$ to $2\pi$ respectively, \cite{kammoun2020asymptotic}. Here, $d_{\text{BS}}=0.5 \lambda$ and $d_{\text{RIS}}=0.5\lambda$ are the inter-antenna separation at the BS and the inter-element separation at the RIS respectively. Each of the elements of $\tilde{\vec{H}}_1$ and $\tilde{\vec{h}}_k^u$ are standard normal RVs. Finally, $\Vec{H}_k$ is defined as $\Vec{H}_k =  \left[\mathbf{h}_{1}^{\mathrm{u}} \ \cdots \ \mathbf{h}_{K}^{\mathrm{u}} \right]$. Similarly, in the current set of simulations, we assume that all the users are data users and hence have $\text{SAR}_{\text{ref},k}=63 \times 10^{-4} \mathrm{~W} / \mathrm{Kg}$ per unit power and $\text{EMF}_{\text{max},k} =0.0029$ for all $k \in \{1,\cdots,K\}$. Similarly, the noise power in dBm is chosen as $\sigma^{2}=-174+10 \log _{10} B$ where the bandwidth for the data user is chosen as $100$ MHz. The maximum transmit power of each user ($p_{\text{max}}$) is chosen as $500$ mW. Throughout the simulations, the value of $L$ to define the convergence condition is chosen as $L=50$. Also, we have chosen the number of bits available for quantization ($B$) to be 3 bits unless mentioned otherwise. \par
\begin{figure}
    \centering
   \includegraphics[scale=0.7]{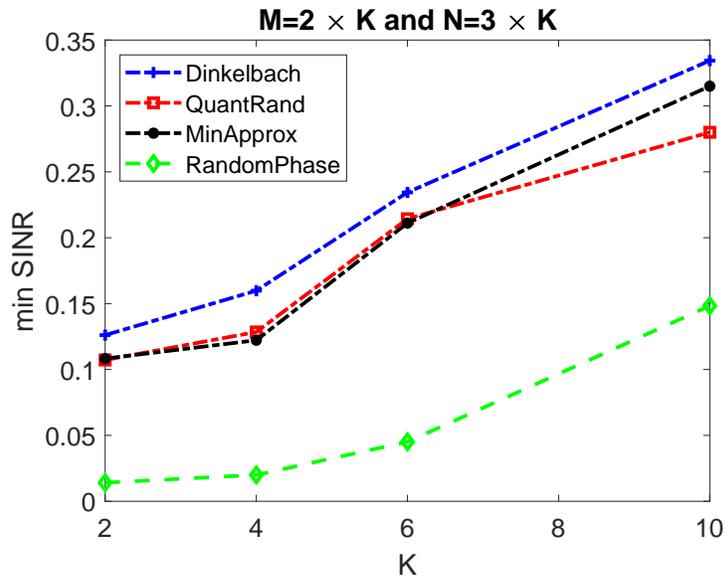}
	\caption{Minimum SINR vs $K$}
	\label{inst_montenum_65_minsinr}
\end{figure}

\begin{figure}
    \centering
	\includegraphics[scale=0.7]{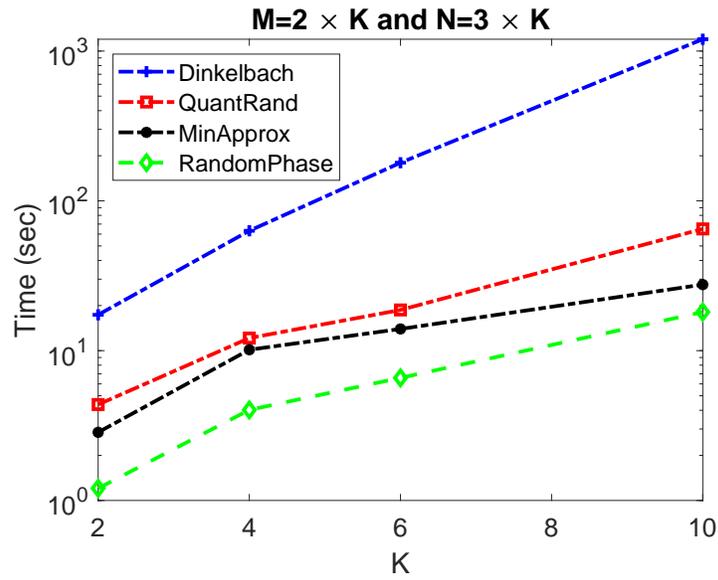}
	\caption{Time for evaluation vs $K$}
	\label{inst_montenum_65_time11}
\end{figure}

 Figure \ref{inst_montenum_65_minsinr} compares the minimum SINR obtained using the alternating optimization technique with different methods proposed for the phase optimization for different values of $K$, $M$ and $N$. All the three curves in Figure \ref{inst_montenum_65_minsinr} perform the beamforming and power allocation according to the solution discussed in Section \ref{bf_opti} and Section \ref{power_opti} respectively. We can observe that the phase shift design using the smooth approximation for the minimum demonstrates the best performance amongst the three schemes compared. The time taken for evaluation for all the three methods in Fig \ref{inst_montenum_65_minsinr}
is shown in Fig \ref{inst_montenum_65_time11}. Here, we can see that the alternating optimization using the Dinkelbach algorithm consumes a large amount of time when compared to the other two schemes.
\par
\begin{figure}
    \centering
    \includegraphics[scale=0.7]{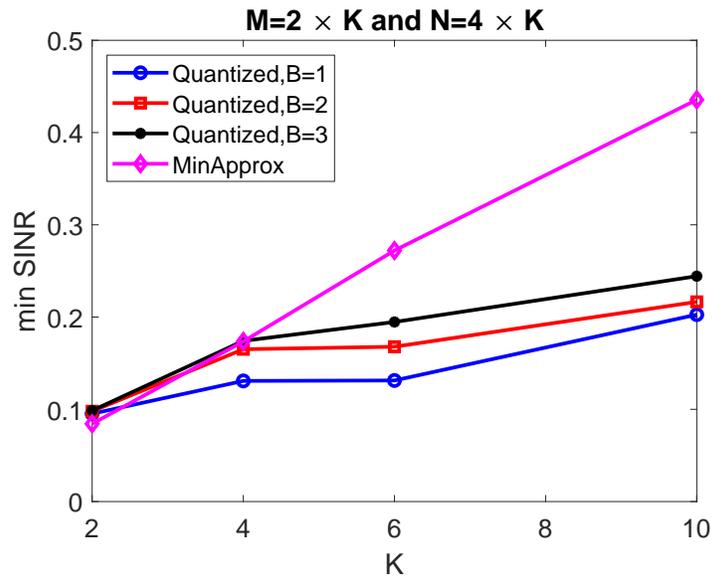}
	\caption{Minimum SINR vs $K$ for different $B$}
	\label{april30_montenum_20_minsinr_num_bits}
\end{figure}

 In Fig \ref{april30_montenum_20_minsinr_num_bits}, we compare the minimum SINR obtained using the quantized phase shift design scheme for different values of $B$. As expected, we can see that the performance improves as the value of $B$ increases.  

\section{Conclusion}
An alternating optimization scheme for finding the optimal receive BF, transmit power, and the RIS phase shifts that maximize the minimum SINR under constraints on the power consumed by each user is proposed. A closed-form expression for the optimal BF vector and a GP-based solution for the power allocation is presented. 
{Algorithms to find the optimal phase shift at the RIS elements assuming both continuous and discrete phase values are presented. 
The proposed methods are compared in terms of the minimum SINR achieved and the time taken for evaluation. Considering a $24$-element RIS, $12$-antenna BS, and $6$ users,  numerical results show that the proposed algorithm achieves close to $300 \%$ gain in terms of minimum SINR compared to a scheme with random RIS phases.}

\begin{appendices}
	\section{Derivative of $\rho_k$}\label{d_rhok_by_phin_appendix}

To evaluate the derivative, 
\begin{align}
 \frac{\partial \rho_k}{\partial \phi_n}   = & {p}_k^* \ \frac{\partial}{\partial \phi_n } Tr{\vec{g}}_k{\vec{g}}_k^H \left( \sum\limits_{i\neq k}^K {p}_i^* {\vec{g}}_i{\vec{g}}_i^H + {\sigma}^2 \vec{I} \right)^{-1} \nonumber  \\  = & {p}_k^*  \frac{\partial}{\partial \phi_n } Tr\left[ \vec{H}_1 \mathbf{R}_{\textrm{R I S}}^{\frac{1}{2}} \vec{\Phi} \vec{R}_{\text{users}}^{\frac{1}{2}} {\vec{h}}_{2,k}{\vec{h}}_{2,k}^H \vec{R}_{\text{users}}^{\frac{1}{2}}\vec{\Phi}^H  \mathbf{R}_{\textrm{R I S}}^{\frac{1}{2}} \vec{H}_1^H \times \right. \nonumber \\  & \left.  \left( \sum\limits_{i\neq k}^K {p}_i^* \vec{H}_1 \mathbf{R}_{\textrm{R I S}}^{\frac{1}{2}} \vec{\Phi} \vec{R}_{\text{users}}^{\frac{1}{2}}{\vec{h}}_{2,i}{\vec{h}}_{2,i}^H  \vec{R}_{\text{users}}^{\frac{1}{2}}   \vec{\Phi}^H\mathbf{R}_{\textrm{R I S}}^{\frac{1}{2}}  \vec{H}_1^H + {\sigma}^2 \vec{I} \right)^{-1} \right]. \nonumber 
\end{align}


Let, $T_1:= \vec{H}_1 \mathbf{R}_{\textrm{R I S}}^{\frac{1}{2}} \vec{\Phi} \vec{R}_{\text{users}}^{\frac{1}{2}} {\vec{h}}_{2,k}{\vec{h}}_{2,k}^H \vec{R}_{\text{users}}^{\frac{1}{2}} \vec{\Phi}^H\mathbf{R}_{\textrm{R I S}}^{\frac{1}{2}}\vec{H}_1^H$ and \\ $T_2:=\left( \sum\limits_{i\neq k}^K {p}_i^* \vec{H}_1 \mathbf{R}_{\textrm{R I S}}^{\frac{1}{2}} \vec{\Phi} \vec{R}_{\text{users}}^{\frac{1}{2}}{\vec{h}}_{2,i}{\vec{h}}_{2,i}^H \vec{R}_{\text{users}}^{\frac{1}{2}} \vec{\Phi}^H  \mathbf{R}_{\textrm{R I S}}^{\frac{1}{2}} \vec{H}_1^H \right.$
	$\left.   + {\sigma}^2 \vec{I} \right)^{-1} $. Then, we make use of the following relation to evaluate the derivative in the last expression: $\frac{\partial K^{-1}}{\partial  p}=-K^{-1} \frac{\partial K}{\partial p} K^{-1}$. Thus, we have, 
\begin{align}
\frac{\partial \rho_k}{\partial \phi_n}  & = p_k^* \frac{\partial}{\partial \phi_n} \text{Tr} \left(T_1 \times T_2 \right) , \nonumber   \\ & = p_k^*\text{Tr} \left(\frac{\partial  T_1}{\partial \phi_n} T_2+T_1\frac{\partial T_2}{\partial \phi_n} \right), \nonumber \\ & =  p_k^*\text{Tr} \left(\frac{\partial  T_1}{\partial \phi_n} T_2-T_1 T_2\frac{\partial \left(T_2 \right)^{-1}}{\partial \phi_n} T_2 \right), \nonumber  \\ & =  p_k^*\text{Tr}\left(\frac{\partial  T_1}{\partial \phi_n} T_2 \right)-p_k^*\text{Tr}\left(T_1 T_2\frac{\partial \left(T_2 \right)^{-1}}{\partial \phi_n} T_2  \right). \nonumber 
\end{align}
	
Next, let us evaluate each of the terms separately. $$\text{Tr}\left(\frac{\partial  T_1}{\partial \phi_n} T_2 \right) = \sum\limits_{j} \sum\limits_{m} \left[T_2 \right]_{m,j} \frac{\partial }{\partial \phi_n} \left[T_1 \right]_{j,m}.$$ 
	Now, we have, 
	\begin{align}
	    \frac{\partial \left[T_1 \right]_{j,m}}{\partial \phi_n}& =\frac{\partial }{\partial \phi_n}\left[\vec{H}_1 \mathbf{R}_{\textrm{R I S}}^{\frac{1}{2}} \vec{\Phi} \underbrace{\vec{R}_{\text{users}}^{\frac{1}{2}} {\vec{h}}_{2,k}{\vec{h}}_{2,k}^H\vec{R}_{\text{users}}^{\frac{1}{2}}}_{\vec{V}_k} \vec{\Phi}^H \mathbf{R}_{\textrm{R I S}}^{\frac{1}{2}}   \vec{H}_1^H  \right]_{j,m}, \nonumber \\
	    & = \left[ \vec{H}_1 \mathbf{R}_{\textrm{R I S}}^{\frac{1}{2}} \vec{\Phi} \vec{V}_k \right]_{j,n} \left[ \left(\vec{H}_1 \mathbf{R}_{\textrm{R I S}}^{\frac{1}{2}}   \right)^H \right]_{n,m}.\nonumber 
	\end{align}

Thus, we have, 
\begin{align}
  \text{Tr}\left(\frac{\partial  T_1}{\partial \phi_n} T_2 \right)  & =\sum\limits_{j} \sum\limits_{m} \left[T_2 \right]_{m,j} \left[ \vec{H}_1 \mathbf{R}_{\textrm{R I S}}^{\frac{1}{2}} \vec{\Phi} \vec{V}_k \right]_{j,n} \left[ \mathbf{R}_{\textrm{R I S}}^{\frac{1}{2}} \vec{H}_1^H \right]_{n,m}  \nonumber  \\ & = \left[\mathbf{R}_{\textrm{R I S}}^{\frac{1}{2}} \vec{H}_1^H  T_2   \vec{H}_1 \mathbf{R}_{\textrm{R I S}}^{\frac{1}{2}} \vec{\Phi} \vec{V}_k\right]_{n,n}. \nonumber 
\end{align} 
Next,  $$\text{Tr} \ T_1 T_2 {\frac{\partial \left(T_2 \right)^{-1}}{\partial \phi_n} } T_2 \sum\limits_{m} \sum\limits_{j} \sum\limits_{\ell} \sum\limits_{s} \left[ T_1\right]_{m,\ell} \left[ T_2\right]_{\ell,j} \left[ T_3\right]_{j,s}  \left[ T_2\right]_{s,m},  $$ where $T_3={\frac{\partial \left(T_2 \right)^{-1}}{\partial \phi_n} }$. We next evaluate $\left[T_3\right]_{j,s}$ as follows, \begin{align}
\frac{\partial \left[\left( T_2 \right)^{-1}\right]_{j,s}}{\partial \phi_n}    == \frac{\partial }{\partial \phi_n}\left[ \sum\limits_{i\neq k}^K {p}_i^* \vec{H}_1 \mathbf{R}_{\textrm{R I S}}^{\frac{1}{2}} \vec{\Phi} \vec{R}_{\text{users}}^{\frac{1}{2}}{\vec{h}}_{2,i}{\vec{h}}_{2,i}^H  \vec{R}_{\text{users}}^{\frac{1}{2}} \vec{\Phi}^H  \mathbf{R}_{\textrm{R I S}}^{\frac{1}{2}}  \vec{H}_1^H + {\sigma}^2 \vec{I} \right]_{j,s} \nonumber 
\end{align}
  Let, $\Vec{V}_{i,k}=\vec{R}_{\text{users}}^{\frac{1}{2}}{\vec{h}}_{2,i}{\vec{h}}_{2,i}^H  \vec{R}_{\text{users}}^{\frac{1}{2}}$. Then, we have, $$\left[T_3\right]_{j,s}=\sum\limits_{i\neq k}^K {p}_i^* \underbrace{\left[ \vec{H}_1 \mathbf{R}_{\textrm{R I S}}^{\frac{1}{2}} \vec{\Phi} \vec{V}_k \right]_{j,n}}_{T_{3}^{1,i}} \underbrace{\left[ \left(\vec{H}_1 \mathbf{R}_{\textrm{R I S}}^{\frac{1}{2}}   \right)^H \right]_{n,s}}_{T_{3}^{2,i}}.$$ Thus, we have, \begin{align}
      \text{Tr} \left(T_1 T_2 T_3 T_2 \right) & =\sum\limits_{i\neq k}^K {p}_i^* \sum\limits_{m} \sum\limits_{j} \sum\limits_{\ell} \sum\limits_{s}\left[ T_1\right]_{m,\ell} \left[ T_2\right]_{\ell,j} \left[ T_3^{1,i}\right]_{j,n} \left[ T_3^{2,i}\right]_{n,s}  \left[ T_2\right]_{s,m} \nonumber \\ & =\sum\limits_{i\neq k}^K {p}_i^* \left[ T_3^{2,i}T_2 T_1 T_2T_3^{1,i} \right]_{n,n}. \nonumber 
  \end{align}
\end{appendices}

\bibliographystyle{IEEEtran}
\bibliography{reference}
\end{document}